\documentclass[11pt,tightenlines,superscriptaddress,floatfixolumn,english,aps,notitlepage,nofootinbib]{revtex4-1}

\usepackage{amsmath}
\usepackage{esint}
\usepackage{graphicx}
\usepackage{hyperref}
\usepackage{color}
\usepackage{xcolor}
\usepackage[utf8]{inputenc} 
\advance\voffset -0.2in

\begin{document}

\title{Elliptic flow coefficients from transverse momentum conservation}

\author{Adam Bzdak}
\email[E-Mail:]{bzdak@fis.agh.edu.pl}
\affiliation{AGH University of Science and Technology,\\ 
Faculty of Physics and Applied Computer Science,
30-059 Krak\'ow, Poland}

\author{Guo-Liang Ma}
\email[E-Mail:]{glma@sinap.ac.cn}
\affiliation{Shanghai Institute of Applied Physics,\\ 
Chinese Academy of Sciences, Shanghai 201800, China}

\begin{abstract}
We calculate the $k$-particle ($k=2,4,6,8$) azimuthal cumulants 
resulting from the conservation of transverse momentum. 
We find that $c_2\{k\}>0$ and depending on the transverse momenta, $c_2\{k\}$ can reach substantial values even for 
a relatively large number of particles. The impact of our results on the understanding 
of the onset of collectivity in small systems is emphasized.
\end{abstract}

\maketitle

\section{Introduction}
\label{sec:introduction}

Relativistic heavy-ion collisions can create an extreme experimental environment with high temperature or high baryon chemical potential that possibly liberates quarks and gluons into a deconfined partonic matter, namely a quark-gluon plasma (QGP)~\cite{Collins:1974ky,Shuryak:1980tp}. The Relativistic Heavy-Ion Collider (RHIC) and the Large Hadron Collider (LHC) have been devoted to investigating the properties of QGP and its transition to hadronic matter. Many experimental results indicate that a strongly interacting partonic matter has been produced in high-energy nucleus-nucleus (A+A) collisions at both RHIC and the LHC~\cite{Adams:2005dq,Adcox:2004mh,Aamodt:2008zz}. One of the most important experimental observations is the collective flow phenomenon, which is believed to arise from the hydrodynamical evolution of QGP, which can convert the initial geometry fluctuations into the final particle momentum anisotropies~\cite{Ollitrault:1992bk,Shuryak:2004cy,Song:2017wtw}. Different experimental methods were invented to measure the coefficients of collective flow  $v_{n}$, e.g., the event plane $v_{n}\{EP\}$ method, two-particle correlations $v_{n}\{2\}$, or multi-particle cumulants $v_{n}\{k\}$ \cite{Borghini:2000sa,Borghini:2001vi,Adler:2002pu,Alt:2003ab}, which show different sensitivities to flow and non-flow effects~\cite{Poskanzer:1998yz,Bilandzic:2010jr}. Hydrodynamical models have successfully reproduced the main features of the measured flow coefficients, see, e.g., Refs.~\cite{Florkowski:book,Teaney:2001av,Song:2007ux,Luzum:2008cw,Bozek:2009dw,Schenke:2010rr}. It indicates that a strongly interactive and collective matter has been produced in A+A collisions.  

Recent experimental results showed long-range in rapidity (i.e., across a large rapidity gap) azimuthal correlations in high-multiplicity proton-proton (p+p) and proton-nucleus (p+A) collisions. 
The extracted $v_{n}$ coefficients are comparable to those measured in A+A collisions~\cite{Khachatryan:2010gv,Abelev:2012ola,Aad:2012gla,Chatrchyan:2013nka}, see also a recent experimental review \cite{Loizides:2016tew}. Similar effects were observed in p+Au, d+Au, and $^{3}$He+Au collisions at RHIC \cite{Adare:2013piz,Adare:2015ctn,McGlinchey:2017esf}, indicating the importance of initial geometry on the observed azimuthal anisotropies. As already envisioned in Refs. \cite{Landau:1953gs,Belenkij:1956cd} these results can be described by hydrodynamical and transport models, see, e.g., Refs.~\cite{Bozek:2011if,Bzdak:2013zma,Shuryak:2013ke,Qin:2013bha,Bozek:2013uha,Ma:2014pva,Bzdak:2014dia,
Koop:2015wea,Shen:2016zpp,Weller:2017tsr,Dusling:2015gta,Song:2017wtw}, indicating that collective flow also exists in small but dense systems. 
The more detailed experimental measurements find that the elliptic flow coefficient is smaller in p+p and p+Pb than in Pb+Pb for a given multiplicity at the LHC energy, but it does not seem to turn off at low multiplicities~\cite{Aaboud:2016yar,Aaboud:2017acw}.\footnote{$v_2\{2\}$ can even increase with decreasing event multiplicity in d+Au collisions at RHIC energies \cite{Aidala:2017ajz}, which is likely due to the non-flow effects, such as the conservation of transverse momentum discussed in this paper.} 
These features seem to challenge our current understanding of collectivity based on the hydrodynamical picture, since it is expected that hydrodynamics should not be applicable to low multiplicity events, where the initial state effects are expected to be well visible, see, e.g.,~\cite{Dumitru:2010iy,Dusling:2013qoz,Skokov:2014tka,Schenke:2015aqa,Schlichting:2016sqo,Kovner:2016jfp,Iancu:2017fzn}. See also Refs. \cite{He:2015hfa,Nagle:2017sjv,Ortiz:2016kpz,Trainor:2017ihu} for other related ideas. 

To shed more light on the low multiplicity {\it collectivity} we will explore the effect of transverse momentum conservation (TMC), which is a well-known azimuthal correlation between all produced particles~\cite{Borghini:2000cm,Borghini:2002mv,Chajecki:2008vg,Chajecki:2008yi,Pratt:2010zn,Bzdak:2010fd}. TMC is an important background in direct flow $v_{1}$ measurements, where its influence has to be corrected especially in peripheral A+A collisions~\cite{ATLAS:2012at,Alt:2003ab}, where $1/N$ corrections due to TMC can be sizable. Therefore, it is essential to explore how TMC influences the elliptic flow coefficients $v_{2}\{k\}$ in small systems, where the multiplicity could be very low and the effect of TMC could be significant.
In particular, the presumed long-range character of TMC complicates the interpretation of the subevent cumulant method proposed recently in Refs. \cite{Jia:2017hbm,Aaboud:2017blb}.

In this paper we calculate 
\begin{equation}
c_{2}\{2\}=\left\langle e^{i2(\phi _{1}-\phi _{2})}\right\rangle ,
\end{equation}%
originating from the conservation of transverse momentum. We also calculate the leading terms of higher order correlation functions \cite{Borghini:2000sa,Borghini:2001vi,Adler:2002pu,Alt:2003ab,Aaboud:2017acw} 
\begin{equation}
c_{2}\{4\}=\left\langle e^{i2(\phi _{1}+\phi _{2}-\phi _{3}-\phi
_{4})}\right\rangle -2\left\langle e^{i2(\phi _{1}-\phi _{2})}\right\rangle
^{2},  \label{eq:v24-def}
\end{equation}%
\begin{eqnarray}
c_{2}\{6\} &=&\left\langle e^{i2(\phi _{1}+\phi _{2}+\phi _{3}-\phi
_{4}-\phi _{5}-\phi _{6})}\right\rangle - \\
&&9\left\langle e^{i2(\phi _{1}-\phi _{2})}\right\rangle \left\langle
e^{i2(\phi _{1}+\phi _{2}-\phi _{3}-\phi _{4})}\right\rangle +12\left\langle
e^{i2(\phi _{1}-\phi _{2})}\right\rangle ^{3},  \notag
\end{eqnarray}
and
\begin{eqnarray}
c_{2}\{8\} &=&\left\langle e^{i2(\phi _{1}+\phi _{2}+\phi _{3}+\phi
_{4}-\phi _{5}-\phi _{6}-\phi _{7}-\phi _{8})}\right\rangle - \\
&&16\left\langle e^{i2(\phi _{1}-\phi _{2})}\right\rangle \left\langle
e^{i2(\phi _{1}+\phi _{2}+\phi _{3}-\phi _{4}-\phi _{5}-\phi
_{6})}\right\rangle -18\left\langle e^{i2(\phi _{1}+\phi _{2}-\phi _{3}-\phi
_{4})}\right\rangle ^{2}+  \notag \\
&&144\left\langle e^{i2(\phi _{1}-\phi _{2})}\right\rangle ^{2}\left\langle
e^{i2(\phi _{1}+\phi _{2}-\phi _{3}-\phi _{4})}\right\rangle
-144\left\langle e^{i2(\phi _{1}-\phi _{2})}\right\rangle ^{4},  \notag
\end{eqnarray}%
which are directly measured at RHIC and the LHC. If $c_2\{k\}$ cumulants are free of non-flow 
effects, then they can be related to the $k$-particle elliptic flow coefficients, $v_{2}\{k\}$, using
\begin{equation}
(v_{2}\{2\})^{2}=c_{2}\{2\},\quad (v_{2}\{4\})^{4}=-c_{2}\{4\},\quad
(v_{2}\{6\})^{6}=\frac{c_{2}\{6\}}{4},\quad (v_{2}\{8\})^{8}=-\frac{%
c_{2}\{8\}}{33}. \label{eq:v2k-c24}
\end{equation}

We find that TMC results in $c_{2}\{k\}>0$ for the calculated $k=2,4,6,8$. In addition, 
$c_{2}\{k\}\sim 1/N^{k}$ for
large $N$, where $N$ is the number of particles subjected to transverse
momentum conservation.

In the next section we present the details of our calculations. Next we
discuss the implications of our results and in the last section we give our
conclusions.

\section{Calculation}
\label{sec:calculation}

In our calculations we assume that transverse momentum conservation is the only source of
correlations between final particles. By $\vec{p}_{i}$ we denote the transverse momentum of the $i$-th particle
emitted in a collision. The $N$ particle transverse momentum distribution
(normalized to unity) $f_{N}$ with imposed transverse momentum conservation
is given by\footnote{$N$ is the total (full phase space) number of particles when the global conservation of transverse momentum is considered. However, as pointed out in Ref. \cite{Pratt:2010zn}, the local (in rapidity) conservation of transverse momentum is likely more physical and thus $N$ can be smaller than the total number of particles.} 
\begin{equation}
f_{N}(\vec{p}_{1},...,\vec{p}_{N})=\frac{1}{A}\delta ^{2}(\vec{p}_{1}+\ldots
+\vec{p}_{N})f(\vec{p}_{1})\cdots f(\vec{p}_{N}),
\end{equation}%
where $f(\vec{p})$ is the single particle transverse momentum distribution
and 
\begin{equation}
A=\int_{F}\delta ^{2}(\vec{p}_{1}+\ldots +\vec{p}_{N})f(\vec{p}_{1})\cdots f(%
\vec{p}_{N})d^{2}\vec{p}_{1}\cdots d^{2}\vec{p}_{N},
\end{equation}%
with the integral taken over the full phase space denoted by $F$. The
details of this calculation can be found in, e.g., 
Refs. \cite{Borghini:2000cm,Chajecki:2008vg,Chajecki:2008yi,Bzdak:2010fd}.

After integrating out all but $k$ momenta we obtain%
\begin{equation}
f_{k}(\vec{p}_{1},...,\vec{p}_{k})=\frac{1}{A}f(\vec{p}_{1})\cdots f(\vec{p}%
_{k})\int_{F}\delta ^{2}(\vec{p}_{1}+\ldots +\vec{p}_{N})f(\vec{p}%
_{k+1})\cdots f(\vec{p}_{N})d^{2}\vec{p}_{k+1}\cdots d^{2}\vec{p}_{N},
\end{equation}%
which, using the central limit theorem, can be approximated by\footnote{See, e.g., Ref. \cite{Begun:2012fd} where the influence of TMC on the particle momentum spectra was calculated exactly, i.e., without using the central limit theorem.}
\begin{equation}
f_{k}(\vec{p}_{1},...,\vec{p}_{k})=f(\vec{p}_{1})\cdots f(\vec{p}_{k})\frac{N%
}{N-k}\exp \left( -\frac{(\vec{p}_{1}+...+\vec{p}_{k})^{2}}{%
(N-k)\left\langle p^{2}\right\rangle _{F}}\right) ,
\end{equation}%
where 
\begin{equation}
\langle p^{2}\rangle _{F}=\frac{\int_{F}p^{2}f(\vec{p})d^{2}\vec{p}}{%
\int_{F}f(\vec{p})d^{2}\vec{p}},
\end{equation}%
with the integration over the full phase space $F$.

\subsection{Two particles}

For two particles we obtain%
\begin{equation}
f_{2}(\vec{p}_{1},\vec{p}_{2})=f(\vec{p}_{1})f(\vec{p}_{2})\frac{N}{N-2}\exp
\left( -\frac{p_{1}^{2}+p_{2}^{2}+2p_{1}p_{2}\cos (\phi _{1}-\phi _{2})}{%
(N-2)\left\langle p^{2}\right\rangle _{F}}\right) ,
\end{equation}%
where $p_{i}=|\vec{p}_{i}|$ and $\phi _{1}-\phi _{2}$ is the azimuthal angle
difference between the two particles.

Using%
\begin{equation}
\left\langle e^{i2(\phi _{1}-\phi _{2})}\right\rangle =\frac{\int_{0}^{2\pi
}f_{2}(\vec{p}_{1},\vec{p}_{2})e^{i2(\phi _{1}-\phi _{2})}d\phi _{1}d\phi
_{2}}{\int_{0}^{2\pi }f_{2}(\vec{p}_{1},\vec{p}_{2})d\phi _{1}d\phi _{2}},
\end{equation}%
we obtain $c_{2}\{2\}$ at a given $p_{1}$ and $p_{2}$
\begin{equation}
c_{2}\{2\}|_{p_{1},p_{2}}=\frac{I_{2}\left( x\right) }{I_{0}\left(
x\right) },\quad x=\frac{2p_{1}p_{2}}{(N-2)\left\langle p^{2}\right\rangle
_{F}},  \label{eq:v22p1p2B}
\end{equation}%
where $I_{k}(x)$ is the modified Bessel function of the the first kind.\footnote{Additionally, 
we have $c_{1}\{2\}=-I_{1}\left( x\right) /I_{0}\left(
x\right) \sim -1/N$ and $c_{3}\{2\}=-I_{3}\left( x\right)
/I_{0}\left( x\right) \sim -1/N^{3}$. }
Expanding Eq. (\ref{eq:v22p1p2B}) in powers of $x$, $I_{2}(x)/I_{0}(x) \approx \frac{x^{2}}{8}$, we obtain 
\begin{equation}
c_{2}\{2\}|_{p_{1},p_{2}}\approx \frac{p_{1}^{2}p_{2}^{2}}{%
2(N-2)^{2}\left\langle p^{2}\right\rangle _{F}^{2}},\text{\qquad }p_{1}p_{2}<%
\frac{1}{2}(N-2)\langle p^{2}\rangle _{F}.  \label{eq:v22p1p2}
\end{equation}

Finally, let us calculate the integrated $c_{2}\{2\}$ over a given
transverse momentum interval. It turns out that 
\begin{equation}
c_{2}\{2\}=\frac{\int dp_{1}dp_{2}p_{1}p_{2}f(p_{1})f(p_{2})\exp
\left( -\frac{p_{1}^{2}+p_{2}^{2}}{(N-2)\left\langle p^{2}\right\rangle _{F}}%
\right) I_{2}\left( \frac{2p_{1}p_{2}}{(N-2)\left\langle p^{2}\right\rangle
_{F}}\right) }{\int dp_{1}dp_{2}p_{1}p_{2}f(p_{1})f(p_{2})\exp \left( -\frac{%
p_{1}^{2}+p_{2}^{2}}{(N-2)\left\langle p^{2}\right\rangle _{F}}\right)
I_{0}\left( \frac{2p_{1}p_{2}}{(N-2)\left\langle p^{2}\right\rangle _{F}}%
\right) },
\end{equation}%
can be very well approximated by%
\begin{equation}
c_{2}\{2\}\simeq \frac{I_{2}\left( \hat{x}\right) }{I_{0}\left( \hat{x}%
\right) },\qquad \hat{x}=\frac{2\sqrt{\left\langle p_{1}^{2}\right\rangle
_{\Omega }\left\langle p_{2}^{2}\right\rangle _{\Omega }}}{\left( N-2\right)
\left\langle p^{2}\right\rangle _{F}},
\end{equation}%
where $\langle p^{2}\rangle _{\Omega }$ is the average of $p^{2}$ in a
measured transverse momentum interval denoted by $\Omega $. For not-too-large 
momenta, $\hat{x}<1$, we obtain%
\begin{equation}
c_{2}\{2\}\approx \frac{\left\langle p_{1}^{2}\right\rangle _{\Omega
}\left\langle p_{2}^{2}\right\rangle _{\Omega }}{2\left( N-2\right)
^{2}\left\langle p^{2}\right\rangle _{F}^{2}}.
\end{equation}

\subsection{Four particles}

For four particles we have 
\begin{equation}
f_{4}(\vec{p}_{1},\ldots ,\vec{p}_{4})=f(\vec{p}_{1})\cdots f(\vec{p}_{4})%
\frac{N}{N-4}\exp \left( -\frac{p_{1}^{2}+p_{2}^{2}+p_{3}^{2}+p_{4}^{2}}{%
(N-4)\left\langle p^{2}\right\rangle _{F}}\right) \exp \left( -\Phi \right) ,
\end{equation}%
where%
\begin{equation}
\Phi =\frac{2}{(N-4)\left\langle p^{2}\right\rangle _{F}}\sum_{i,j=1;\text{ }%
i<j}^{4}p_{i}p_{j}\cos (\phi _{i}-\phi _{j}).  \label{eq:big-phi}
\end{equation}

To calculate the four-particle correlator at a given transverse momenta $%
p_{1}$, $p_{2}$, $p_{3}$ and $p_{4}$ 
\begin{equation}
\left\langle e^{i2(\phi _{1}+\phi _{2}-\phi _{3}-\phi _{4})}\right\rangle
|_{p_{1},p_{2},p_{3},p_{4}}=\frac{\int_{0}^{2\pi }d\phi _{1}\cdots d\phi
_{4}\exp \left( -\Phi \right) e^{i2(\phi _{1}+\phi _{2}-\phi _{3}-\phi _{4})}%
}{\int_{0}^{2\pi }d\phi _{1}\cdots d\phi _{4}\exp \left( -\Phi \right) }, \label{eq:big-phi-4}
\end{equation}%
we expand $\exp \left( -\Phi \right) $ in $\Phi $. In the numerator, denoted
by $U$, the first non-vanishing term is given by $\Phi ^{4}/24$ which gives%
\begin{equation}
U\approx 24\pi ^{4}\frac{(p_{1}p_{2}p_{3}p_{4})^{2}}{(N-4)^{4}\left\langle
p^{2}\right\rangle _{F}^{4}},
\end{equation}%
and the remaining terms are suppressed by the higher powers of $N$. In the
denominator, denoted by $D$, we simply have $D\approx 16\pi ^{4}$ resulting
in%
\begin{equation}
\left\langle e^{i2(\phi _{1}+\phi _{2}-\phi _{3}-\phi _{4})}\right\rangle
|_{p_{1},p_{2},p_{3},p_{4}}\approx \frac{3}{2}\frac{%
(p_{1}p_{2}p_{3}p_{4})^{2}}{(N-4)^{4}\left\langle p^{2}\right\rangle _{F}^{4}%
},
\end{equation}%
and using Eq. (\ref{eq:v24-def}) we find 
\begin{equation}
c_{2}\{4\}|_{p_{1},p_{2},p_{3},p_{4}}\approx \frac{%
(p_{1}p_{2}p_{3}p_{4})^{2}}{(N-4)^{4}\left\langle p^{2}\right\rangle _{F}^{4}%
}.  \label{eq:v24p1p4}
\end{equation}
To obtain this result we replaced $N-2$ by $N-4$ in Eq. (\ref{eq:v22p1p2}).

\subsection{Six and eight particles}

Performing analogous calculations we obtain\footnote{%
In this case, the first non-vanishing terms in the expressions analogous to
Eqs. (\ref{eq:big-phi}) and (\ref{eq:big-phi-4}) are given by $\Phi ^{6}/6!$ and $\Phi ^{8}/8!$ for
six and eight particles, respectively.}%
\begin{equation}
\left\langle e^{i2(\phi _{1}+\phi _{2}+\phi _{3}-\phi _{4}-\phi _{5}-\phi
_{6})}\right\rangle |_{p_{1},...,p_{6}}\approx \frac{45}{4}\frac{%
(p_{1}p_{2}p_{3}p_{4}p_{5}p_{6})^{2}}{(N-6)^{6}\left\langle
p^{2}\right\rangle _{F}^{6}},
\end{equation}%
and%
\begin{equation}
\left\langle e^{i2(\phi _{1}+\phi _{2}+\phi _{3}+\phi _{4}-\phi _{5}-\phi
_{6}-\phi _{7}-\phi _{8})}\right\rangle |_{p_{1},...,p_{8}}\approx \frac{315%
}{2}\frac{(p_{1}p_{2}p_{3}p_{4}p_{5}p_{6}p_{7}p_{8})^{2}}{%
(N-8)^{8}\left\langle p^{2}\right\rangle _{F}^{8}},
\end{equation}%
resulting in%
\begin{equation}
\frac{1}{4} c_{2}\{6\}|_{p_{1},...,p_{6}}\approx \frac{3}{2}\frac{%
(p_{1}p_{2}p_{3}p_{4}p_{5}p_{6})^{2}}{(N-6)^{6}\left\langle
p^{2}\right\rangle _{F}^{6}},  \label{eq:v26p1p6}
\end{equation}%
and 
\begin{equation}
\frac{1}{33}c_{2}\{8\}|_{p_{1},...,p_{8}}\approx \frac{24}{11}\frac{%
(p_{1}p_{2}p_{3}p_{4}p_{5}p_{6}p_{7}p_{8})^{2}}{(N-8)^{8}\left\langle
p^{2}\right\rangle _{F}^{8}}.  \label{eq:v28p1p8}
\end{equation}

\section{Results}
\label{sec:results}

In this section we plot $c_{2}\{k\}$ for different ranges of
transverse momenta. We assume%
\begin{equation}
f(p)\propto \exp \left( -p/T\right) ,
\end{equation}%
with $T=0.25$ GeV. In this case $\langle p\rangle _{F}=2T$ and $%
\langle p^{2}\rangle _{F}=6T^{2}$. 

\begin{figure}[t]
\begin{center}
\includegraphics[scale=0.4]{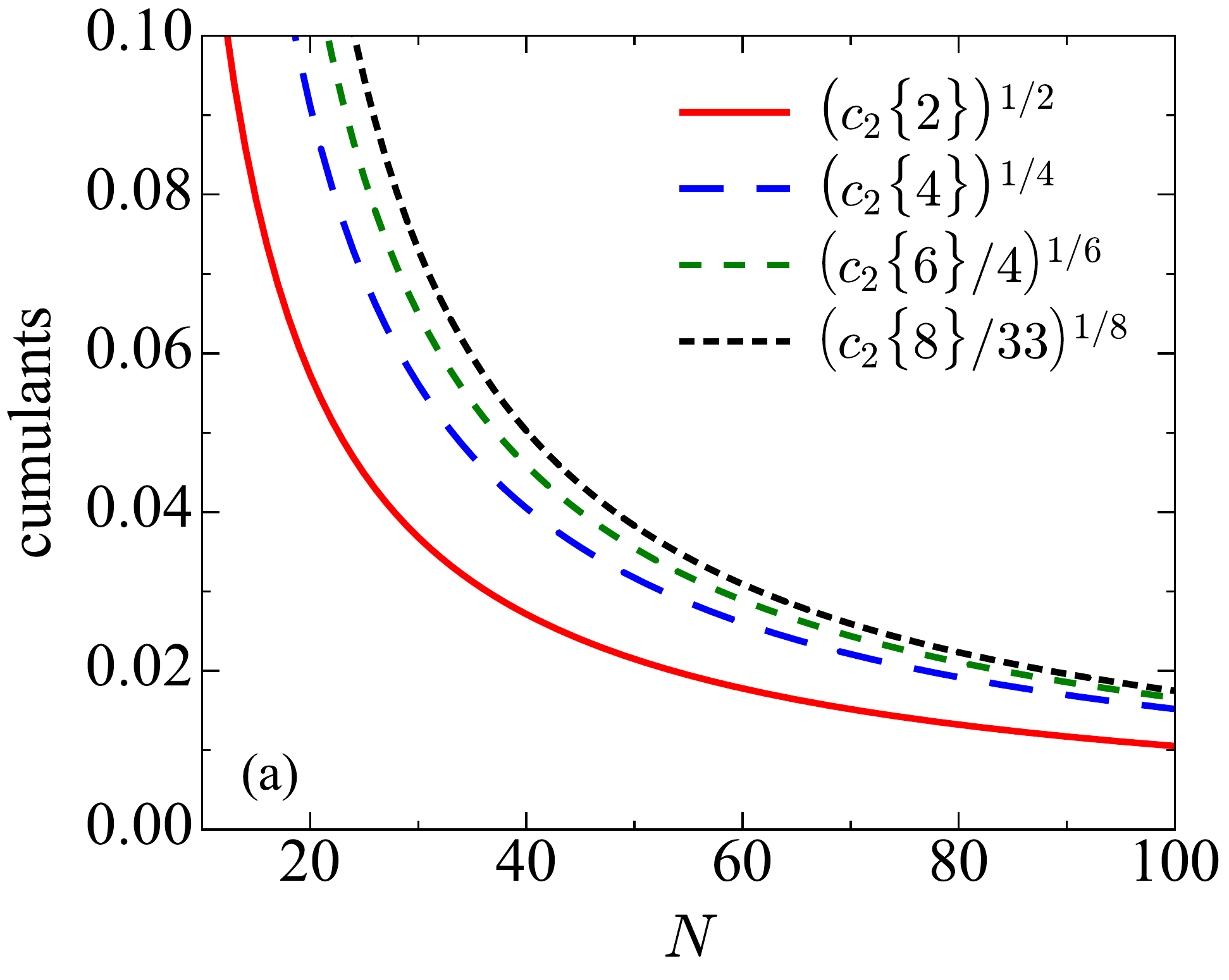}\hspace{5mm} %
\includegraphics[scale=0.4]{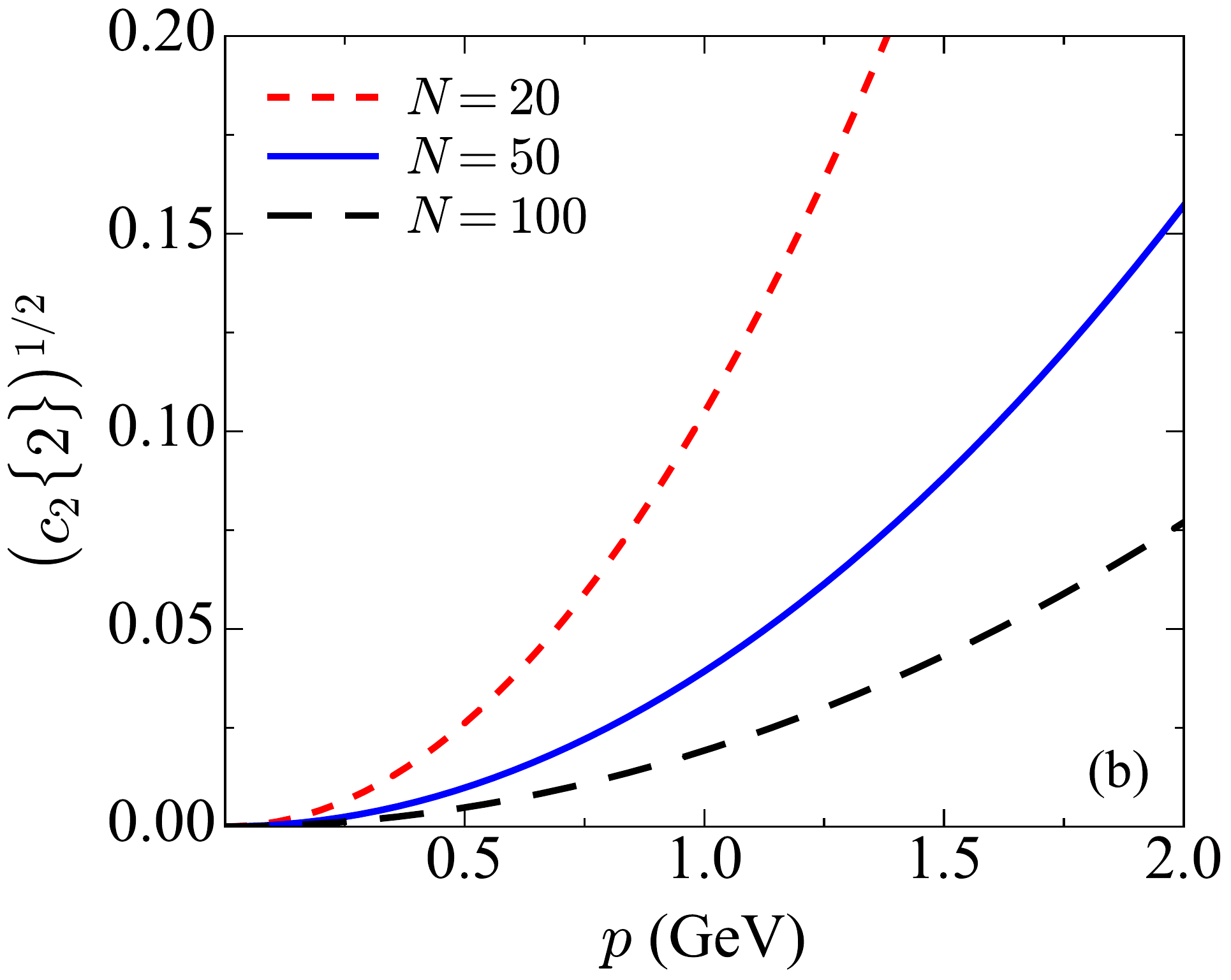}
\end{center}
\par
\vspace{-5mm}
\caption{(a)  
The multi-particle azimuthal cumulants $(c_{2}\{2\})^{1/2},$ $(c_{2}\{4\})^{1/4},$ $(c_{2}\{6\}/4)^{1/6}$, 
and $(c_{2}\{8\}/33)^{1/8}$ as a
function of the number of particles, $N$. (b) $(c_{2}\{2\})^{1/2}$ as function of transverse momentum, $p$, 
for different values of $N=20,$ $50$, and $100$.}
\label{fig:1}
\end{figure}

In the left panel of Fig. \ref{fig:1} we present $(c_{2}\{2\})^{1/2},$ $(c_{2}\{4\})^{1/4},$ $(c_{2}\{6\}/4)^{1/6}$, 
and $(c_{2}\{8\}/33)^{1/8}$ as a function of the number of particles, $N$. In
this calculation all transverse momenta $p_{i}$ from Eqs. (\ref{eq:v22p1p2},%
\ref{eq:v24p1p4},\ref{eq:v26p1p6},\ref{eq:v28p1p8}) are equal to $%
\langle p^{2}\rangle _{\Omega }^{1/2}$, where $\Omega $ is the
transverse momentum range of $p>0.3$ GeV, typically assumed in
experimental measurements. The cumulants originating from transverse momentum conservation are basically inversely proportional to $N$, which reflects an intrinsic feature of TMC. They absolute magnitudes increase with the order of cumulant due to the larger coefficients present in higher order cumulants, see Eqs. (\ref{eq:v22p1p2},\ref{eq:v24p1p4},\ref{eq:v26p1p6},\ref{eq:v28p1p8}). It is interesting that their signs are consistent with the recent experimental measurement of $c_2\{k\}$, where for the lowest multiplicity events $c_2\{k\}>0$ \cite{Aaboud:2017acw}.

In the right panel of Fig. \ref{fig:1} we present 
$(c_{2}\{2\})^{1/2}$ as function of $p=p_{1}$ and $p_{2}=\langle p^{2}\rangle _{\Omega }^{1/2}$ for 
different values of $N=20,$ $50$, and $100$.\footnote{To be precise, we plot $c_2\{2\}(p,p_2) / \sqrt{c_2\{2\}(p_2,p_2)}$, which for Eq. (\ref{eq:v22p1p2}) is equivalent to $\sqrt{c_2\{2\}(p,p)}$.}
As expected, the influence of TMC on $(c_{2}\{2\})^{1/2}$ is more significant for particles with higher momenta (parabolic dependence) and smaller number of particles $N$.

In addition, we calculated $c_2\{2\}$ using the Monte Carlo methods and obtained practically identical result to this shown in the left panel of Fig. \ref{fig:1}. The purpose of this exercise was to verify that the central limit theorem, assumed in our analytical calculations, does not seriously effect the final results.

\section{Conclusions}

In this paper we calculated analytically the multi-particle elliptic flow cumulants originating from the conservation of transverse momentum.\footnote{In our calculations we assumed that the conservation of transverse momentum is the only source of correlations between produced particles. It would be interesting to couple this mechanism with correlations originating from hydrodynamic flow. This should naturally explain the sign change of $c_2\{4\}$ and $c_2\{8\}$ as a function of the number of produced particles. This problem is currently under our investigation.} 
We demonstrated that $c_2\{k\}>0$ ($k=2,4,6,8$)  
in qualitative agreement with the lowest multiplicity data in p+p, p+A and A+A collisions. As expected, the TMC contribution is found to be more significant for particles with higher transverse momenta and for systems with smaller number of particles (due to an intrinsic $1/N$ effect). Depending on the transverse momenta, $(c_2\{k\})^{1/k}$ 
can reach substantial values, of the order of a few percentages, even for a relatively large number of particles. This effect should be taken into account when interpreting the long-range azimuthal correlations in small systems currently measured at RHIC and the LHC.

\bigskip

\vspace{\baselineskip} 
\noindent\textbf{Acknowledgments} 
\newline
{} 
We thank Piotr Bo\.zek for useful comments. 
A.B. is partially supported by the Faculty of Physics and Applied Computer Science AGH UST statutory tasks within subsidy of Ministry of Science and Higher Education, and by the National Science Centre, Grant No. DEC-2014/15/B/ST2/00175.
G.-L.M. is supported by the Major State Basic Research Development Program in China under Grant No. 2014CB845404, the National Natural Science Foundation of China under Grants No. 11522547, 11375251, and 11421505.

\end{document}